\documentclass[12pt]{article}

\usepackage{amssymb}

\date{}

\usepackage{color}

\begin{document}
 \title{{\bf Variational Methods in AdS/CFT}}
 \author{Tom\'{a}s Andrade, M\'{a}ximo Ba\~{n}ados and Francisco Rojas \\  Departamento de F\'{\i}sica,\\
 P. Universidad Cat\'{o}lica de Chile, Casilla 306, Santiago 22, Chile. \\
 {\tt tiandrad@uc.cl, maxbanados@fis.puc.cl,
    frojasf@uc.cl}
}

 \maketitle

\begin{abstract}
We prove that the $AdS/CFT$ calculation of 1-point functions can
be drastically simplified by using variational arguments. We give
a simple proof, valid for any theory that can be derived
from a Lagrangian, that the large radius divergencies in 1-point
functions can always be renormalized away (at least in the
semiclassical approximation).  The renormalized 1-point functions
then follow by a simple variational problem involving only finite
quantities.  Several examples, a massive scalar, gravity, and
renormalization flows, are discussed. Our results are general and
can thus be used for dualities beyond $AdS/CFT$.
\end{abstract}

\section{Introduction and motivation}

In the AdS/CFT correspondence \cite{Maldacena,GKP,Witten}, the key object to be calculated is the generating functional $W[J]$ of connected diagrams associated to an operator ${\cal O}$. According to the proposal put forward in \cite{Maldacena,GKP,Witten}, for each operator ${\cal O}$ there exists a dual bulk field $\Phi$ with
action $I[\Phi]$ such that
\begin{equation}\label{fW}
e^{-W[J]}  = \int_{AdS} D\Phi\, e^{-I[\Phi]}
\end{equation}
The integral in the right hand side depends on the source $J$
through boundary conditions. For a generic asymptotic
expansion\footnote{\label{f1} $\sigma_i$ are typically rational
numbers, and $\sigma_0$ is associated to the conformal dimension
of the dual operator. $\rho $ is the radial coordinate  $\rho
\rightarrow 0$ defining the boundary. There may be logarithmic
terms at some order which are relevant for the structure of the
solution and for the anomalies. At this point, we only need the
existence of some series.},
\begin{equation}\label{bc}
\Phi(x,\rho) \rightarrow \rho^{\sigma_0} \varphi_0(x) +
\rho^{\sigma_1}\varphi_1(x) + \rho^{\sigma_2}\varphi_2(x) + \cdots
, \ \ \ \ \ \ \  \rho\rightarrow 0.
\end{equation}
where $\sigma_0 < \sigma_1 < \sigma_2..$. The leading coefficient
$\varphi_0(x)$ is identified with the source (see
\cite{Klebanov-Witten,Witten01} for other cases),
\begin{equation}\label{source}
J(x) = \varphi_0(x),
\end{equation}
and the path integral in (\ref{fW}) is computed with $\varphi_0(x)$ fixed.

The calculation of the integral in (\ref{fW}) is as difficult, if not more, as the CFT calculation of $W[J]$. What makes AdS/CFT useful is the fact that the correspondence is strong/weak. This means that the semiclassical approximation of (\ref{fW}),
\begin{equation}\label{W}
W[J] \ \simeq \ \left. I\,\left[\Phi(\varphi_0)\right]\  \right |_{on-shell},
\end{equation}
((\ref{source}) is understood) gives strongly coupled results for $W[J]$.

Now, the approximation (\ref{W}) is valid {\it provided} the
action $I[\Phi]$ is stationary under variations with
$\varphi_0(x)$ fixed, that is, the variation of the action must satisfy,
\begin{equation}\label{var}
\delta I[\Phi] = \int_{M}{\cal E}(L)\,\delta\phi \ + \ \int_{\partial M} A\, \delta \varphi_{0}.
\end{equation}
where $A$ is some coefficient, and ${\cal E}(L)$ are the
Euler-Lagrange equations.  If the action $I[\Phi]$ fails to
satisfy (\ref{var}), then the semiclassical approximation is not
valid and, of course, (\ref{W}) is not valid either\footnote{It
sometimes argued that the AdS/CFT prescription
is incomplete because boundary terms can be added to the action
without affecting the equations of motion. This point of view is
incorrect. The action functional $I[\Phi]$ is uniquely defined, up
to trivial scheme dependent terms, once the source $\varphi_0$ is
identified.}.

The importance of the variational problem in AdS/CFT calculations
was stressed in \cite{Henneaux}, where ambiguities in the
holographic description of Dirac spinors \cite{Henningson-Sfetsos}
were fixed. Even for the simple example of massive scalars,
problems with the Ward identities \cite{Freedman-} were solved by
a proper analysis of the variational
problem\cite{Klebanov-Witten,Muck-Viswanathan,HR}. Since then, the
analysis of boundary terms in AdS/CFT, specially for scalar
fields, has received much attention
\cite{HR,Arutyunov-Frolov,Arutyunov-Frolov-Theisen,Minces-Riveles,Braga-Boschi}.
More recently, this problem has revived in the context of multi
trace deformations, e.g.,
\cite{Nolland,Muck,Hatmann-Rastelli,Berkooz-Sever-,Papadimitriou,Marolf-Ross,Minces}.

Besides being stationary, the action $I[\Phi]$, and in particular
the coefficient $A$ in Eq. (\ref{var}), must be finite, otherwise
the equality (\ref{W}) would not be very useful. Since the
background has infinite volume (normally AdS space, but other
backgrounds are also relevant for realistic dualities)
divergencies do arise and they need to be renormalized. This is
done by standard technics. One first introduces a regulator and
then subtracts divergent terms. As first noticed in
\cite{Freedman-,Muck-Viswanathan}, it is necessary to first
formulate a proper Dirichlet problem at at fixed value of the
regulator, and take the limit at the very end. A general
procedure, termed Holographic Renormalization, was put forward in
\cite{HR}, and a Hamiltonian approach was developed in
\cite{HRnew}. In particular, in \cite{HRnew}, general formulas for Asymptotically locally AdS spacetimes have been displayed.  Dimensional regularization is also available with equivalent results \cite{Imbimbo,ST1,ST2}. A
Hamilton-Jacobi method, exploiting the gravitational Hamiltonian
constraint, is also available \cite{deBoer-Verlinde2}.

~

The goal of this paper is to show that variational arguments
can provide simplifications in explicit calculations. Variational approaches have been considered previously in the literature (see \cite{PS} and references therein), including the issue of global charges.  Our goal here is to put forward a novel and simple
approach to deal with ``renormalization problem". We
prove that the infinities can always be renormalized away, for any theory,
at least in the semiclassical approximation (\ref{W}). In particular, we prove that all
divergencies appearing in the variation of the action can be
subtracted without interfering with the finite parts. This means
that the calculation of 1-point functions $\langle {\cal O}
\rangle_J$ (in the presence of sources) can be done via a simple
finite variational problem and divergencies can be ignored
completely\footnote{Of course divergencies contain important
information and in some cases it is preferably to keep them, e.g.,
\cite{Henningson-Skenderis}. But, as far as the semiclassical
1-point function is concerned, we shall see that they can be
ignored completely.}.

Furthermore, our procedure does not make use of the AdS structure
explicitly. Thus, in principle, it could be useful in more
realistic gauge theory/gravity correspondences (see
\cite{Berg-Haack-Muck} for a recent discussion) where explicit
calculations have been hampered by technical difficulties
associated with the renormalization problem. Particular
applications of the method described here have appeared in
\cite{BST,BTOM}.

\section{The variational problem and 1-point functions}

In this section we prove the main result of the paper, namely, that 1-point functions (with non-zero sources) can be computed in AdS/CFT {\it without} calculating the divergent terms.

Consider the general action (see Sec. \ref{HOrder} for a more general action)
\begin{equation}\label{I}
I[\Phi] = \int_\epsilon d\rho \int_{M} d^dx \, L(\Phi,\Phi',\partial_i\Phi) + \int_{\rho=\epsilon} d^dx\,B.
\end{equation}
with associated the Euler-Lagrange equations
\begin{equation}\label{EL}
\left( {\partial L \over \partial\Phi'} \right)' + \partial_i \left( {\partial L \over \partial_i\Phi} \right) = {\partial L \over\partial \Phi}.
\end{equation}
The action is defined on a manifold with asymptotic topology
$\Re\times M$. $\Phi$ denotes collectively the set of fields under
consideration. $\rho$ is a radial coordinate and the boundary is
located at $\rho\rightarrow 0$. We follow closely the notations
introduced in \cite{HR}. The parameter $\epsilon$ is a cutoff that
regularizes the eventual IR divergencies at $\rho\rightarrow 0$.
The primes denote derivatives with respect to $\rho$, which we
have separated from the other derivatives only for illustrative
purposes.

The holographic calculation of field theory correlators dual to an
action $I$ has three different faces. The dictionary problem
establishes the relationship between bulk fields and dual
operators. This is where most of the AdS/CFT physics resides. The
renormalization and Dirichlet problems, namely, to find $B$ to
make the  bulk action finite and well-defined for the given
boundary conditions. Finally, the fluctuation problem involves
imposing boundary conditions in the deep interior and find
non-local relations among the boundary data.  In this paper we
shall only deal with the second issue, namely, the renormalization
problem. We refer to the large literature for the other problems.

~

As usual, we formulate the holographic renormalization problem as
follows: We seek a boundary term $B$ such that the on shell
variation of the action (\ref{I}) satisfies
\begin{eqnarray}
\delta I[\Phi] &=& \lim_{\epsilon\rightarrow 0}\int_{\rho=\epsilon} d^dx \, \left( {\partial L \over\partial \Phi'} \delta \Phi +  \delta B \right)  \label{goal0}\\
&=& \int d^dx\, A \, \delta \varphi_0\label{goal}
\end{eqnarray}
where $A$ is a {\it finite} coefficient, and $\varphi_0$ is the
source  for the given problem.  If (\ref{goal0}) can be written in
the form (\ref{goal}), then, according to \cite{GKP,Witten}, the
problem is solved and the vev of the dual operator is
\begin{equation}\label{O=A}
\langle {\cal O}  \rangle = A
\end{equation}
It is clear from the transition from (\ref{goal0}) to (\ref{goal})
that the boundary term $B$ is playing two roles. On the one hand
it must ensure that the Dirichlet problem is well defined, namely,
$\delta I \sim \int A \delta\varphi_0$. On the other, we demand
$A$ in (\ref{goal}) to be finite, and hence $B$ must remove the
divergencies coming from the bulk. We show here that these two
problems can in fact be analyzed separately.

\subsection{The renormalization problem. }
\label{renor}

Let us first deal with renormalization problem. We prove here that
all divergencies in ${\partial L \over\partial \Phi'} \delta \Phi$
can be written as total variations, and hence be canceled by $B$.
This result does not depend on the explicit form of the series
(\ref{bc}). To fix the ideas, we expand both $\Phi$ and
$\delta\Phi$ in the form (see footnote \ref{f1}),
\begin{eqnarray}\label{FG}
\Phi &=& \rho^{\sigma} \left( \varphi_{0} + \rho \varphi_{1} + \rho^2 \varphi_{2} + \cdots \right) \nonumber\\
\delta\Phi &=& \rho^{\sigma} \left(\delta \varphi_{0} + \rho \delta\varphi_{1} + \rho^2 \delta\varphi_{2} + \cdots \right).
\end{eqnarray}
and insert them in ${\partial L \over\partial \Phi'} \delta \Phi$.
We find an expansion of the form,
\begin{equation}\label{S}
{\partial L \over\partial \Phi'} \delta \Phi =  \sum_{n=-K}^{\infty} \rho^n\,  C_n(\varphi_i,\delta\varphi_j).
\end{equation}
Here, $\varphi_i$ are the coefficients appearing in the series (\ref{FG}). $K$ is some positive number and represent the fact that, generically, (\ref{S}) contains divergent pieces.  There may be $\log$'s and other types of functions of $\rho$.  As it will be clear in a moment, this will not be relevant for the analysis that follows.

We are now going to prove the following general result. All coefficients $C_n(\varphi,\delta\varphi)$, {\it except the zero mode $C_0$}, can be written as a total variation. This means, there exists local functions $D_n(\varphi_i)$ such that,
\begin{equation}\label{Dn}
C_n(\varphi_i,\delta\varphi_j) = \delta D_n(\varphi_i), \ \ \ \ \  \forall\, n\neq 0.
\end{equation}
To this end, consider the following equality,
\begin{eqnarray}
\left( {\partial L \over\partial \Phi'} \delta\Phi \right)' &=& \left( {\partial L \over\partial \Phi'}\right)'  \delta\Phi + {\partial L \over\partial \Phi'} \delta\Phi' \nonumber\\
 &=& \left( {\partial L \over\partial \Phi} - \partial_i \left( {\partial L \over \partial_i\Phi} \right) \right)\delta \Phi  +  {\partial L \over\partial \Phi'} \delta\Phi' \nonumber\\
 &=& {\partial L \over\partial \Phi}\delta\Phi  + \left( {\partial L \over \partial_i\Phi} \right) \delta(\partial_i\Phi)  +  {\partial L \over\partial \Phi'} \delta\Phi'  - \partial_i\left( {\partial L \over \partial_i\Phi} \delta\Phi\right)  \nonumber\\
 &=& \delta L \label{dL}
\end{eqnarray}
In the second line we have used the equations (\ref{EL}),
and in the last line we have used that $L =
L(\Phi,\Phi',\partial_i\Phi)$. We have also discarded a total
derivative term which, under the integral sign, $\int d^dx$, will
not contribute.

Taking the radial derivative of  (\ref{S}) at both sides and
comparing with (\ref{dL}) we conclude that (\ref{Dn}) is in fact
true, and that the functions $D_n$ are the Taylor coefficients of
the Lagrangian $L$. Since divergencies are contained in
$\rho-$dependent contributions, $1/\rho^n$,  $\log\rho$, etc, we
conclude that they can always be subtracted by choosing $B$
appropriately. See sections \ref{toy} and \ref{gravity} for
explicit examples.

Note also that neither the form of the series (\ref{FG}) nor
(\ref{S}) are really relevant. The only important point is to
distinguish between the $\rho-$dependent parts of ${\partial L
\over\partial \Phi'} \delta\Phi$, and its zero mode. According to
the above result, all $\rho-$ dependent parts are total
variations, while no information can be extracted on the zero
mode. This is a completely general result \footnote{The formula
(\ref{dL}) may also be relevant for an off-shell formulation of
the action, and the finiteness of conserved charges. It is known
that when turning on non-standard modes of some fields,
divergencies in the charges are canceled after properly
incorporating the back reaction into the boundary conditions
\cite{HMTZ,BST2}. Presumably, these cancelations can be understood
as special cases of (\ref{dL}). We thank M. Henneaux for a
discussion on this point.}  valid also for higher order theories
(see Sec. \ref{HOrder}).

Summarizing, when computing the on-shell variation of the action
(\ref{I}) one can simply ignore all divergent terms and keep only
the zero mode (in the $\rho$ expansion). The above result is
general enough to ensure the existence of $B$ to subtract all
divergencies and this is all one needs at this point.  The 1-point
functions are finite and only depend on the zero modes of
${\partial L \over\partial \Phi'} \delta\Phi$ and $B$. We now turn
into their analysis.

\subsection{The variational problem}

Having discarded all divergencies by a choice of $B$, the problem (\ref{goal}) is turned into a restricted and much simpler problem, namely, to find $B_0$ such that,
\begin{eqnarray}\label{goal2}
\delta I &=& \int d^dx\, \left( \left. {\partial L \over\partial \Phi'} \delta\Phi \right|_{\mbox{zero mode}}  + \delta B_0 \right), \nonumber\\
&=& \int d^dx\, A\, \delta \varphi_0,
 \end{eqnarray}
where $B_0$ is now the zero mode part of $B$. The finite
coefficient $A$ equals to the dual vev  $\langle {\cal O}  \rangle
= A$. This problem does not have a universal solution and, in
fact, $B_0$ may not even exist at all unless certain integrability
condition is fulfilled.

It is worth noting that $A$ may contain purely local pieces, that
cannot be eliminated unless they are integrable, as in the case of
gravity (see Sec. \ref{CBflow}).

We would like to end this paragraph remarking that this procedure
to deal with the finite part of the on-shell variation of the
action has already appeared in the literature in various examples,
e.g.,
\cite{Arutyunov-Frolov,Arutyunov-Frolov-Theisen,Minces-Riveles,Braga-Boschi,Marolf-Ross,BST,BTOM}. General formulae for asymptotically locally AdS spacetimes are also known \cite{PS}.

\section{Examples}
\label{Examples}

We display in this section some examples and applications of the
variational approach to the holographic calculation of
correlators.  We consider here a toy model, a massive scalar
field, and gravity.

\subsection{Toy model and comparison with Holographic Renormalization} \label{toy}

Let us illustrate the ideas presented so far with a simple
example. Consider the action for a toy model in $0+1$ dimensions,
\begin{equation}
I = {1 \over 2} \int d\rho \left[ {2 \over \rho} \Phi'^2 - {2\kappa \over \rho^2} \Phi^2 - {3 \over 2\rho^3} \Phi^2 \right] + B
\end{equation}
This system has the same features of a scalar field on $AdS_4$
with dual conformal dimension $\Delta=3$, $\kappa$ playing the
role of the Laplacian.  Setting $\Phi = \rho^{1/2} \varphi(\rho)$,
the equation of motion become $\rho \varphi'' + \kappa\varphi=0$,
which has the asymptotic series,
\begin{equation}\label{FGtoy}
\varphi = \varphi_0 + \rho (\varphi_1 + \psi\log\rho) + \cdots
\end{equation}
with $\varphi_0$ and $\varphi_1$ arbitrary while
$\psi=-\kappa\varphi_0$. The exact solution exists in terms of
Bessel functions. The on-shell variation of the action becomes,
\begin{eqnarray}
\delta I &=& \left.{2 \over \rho} \Phi'\delta \Phi \right|_{\rho=\epsilon}+\delta B , \\
   &=&\left[  \delta \varphi_0 \varphi_1 +  \varphi_0 \delta \varphi_1  + 2(\varphi_1+\psi)\delta \varphi_0 + \delta B_0 \right]+ \left[{1 \over\epsilon }\varphi_0\delta \varphi_0 +(\varphi_0 \delta \psi + 3\psi\delta \varphi_0) \log\epsilon+ \delta B(\epsilon)\right] \label{toy2} .
\end{eqnarray}
The first square brackets corresponds to the variation problem and
contain only finite ``zero mode" terms. The second bracket
contains the divergencies, and represents the ``finiteness"
problem. We have separated $B$ in two pieces $B(\epsilon)+B_0$
stressing its double role.

We already know the divergencies can be subtracted by $B(\epsilon)$. But let
us check this explicitly in this example. The second bracket contains $\psi\delta\varphi_0$ which is not an exact variation.
However, using the equation $\psi=-\kappa\varphi_0$ it can be written as  $\delta \left[{\varphi_0^2  \over 2\epsilon} - 2\kappa \varphi_0^2\log\epsilon +
B(\epsilon) \right] $. This is then removed by choosing $B(\epsilon) =  -{\varphi_0^2
\over 2\epsilon} + 2\kappa \varphi_0^2\log\epsilon$. A less trivial example, gravity in four dimensions, is discussed in Sec. \ref{gravity}.

Having discarded the divergencies, we end up with the finite
problem, which is the relevant one for the 1-point function.  The
finite pieces in (\ref{toy2}) are almost in the form (\ref{goal}),
except for the term $ \varphi_0 \delta\varphi_1$. Since the
asymptotic equations do not fix $\varphi_1$ in terms of
$\varphi_0$ we cannot express the variation  $\delta\varphi_1$ in
terms of $\delta\varphi_0$.  At this point $B_0$ enters into the game.
We choose $B_0 = -\varphi_1\varphi_0$ and find,
\begin{equation}\label{Otoy}
\delta I = 2(\varphi_1+\psi)\delta \varphi_0 \ \ \ \ \ \
\Rightarrow \ \ \ \ \  \langle {\cal O}  \rangle =
2(\varphi_1+\psi).
\end{equation}

Two comments are in order. First, the choices made for
$B(\epsilon)$ and $B_0$ are unique, up to terms local in the
source $\varphi_0$, which of course cannot be fixed and lead to
contact terms. Second, the result (\ref{Otoy}) is in complete
agreement with the method described in \cite{HR}.  As a final
point we note that
\begin{equation}\label{check}
B(\epsilon) + B_0 = -{1\over 2\epsilon^2}\,\Phi^2 + {2\kappa \log\epsilon \over \epsilon}\, \Phi^2 + {\cal O}(\epsilon)
\end{equation}
The right hand side of this equation is exactly the counterterm
action $C[\Phi]$ that one would have computed following \cite{HR}.

\subsection{Single massive scalar field} \label{massive scalar}

Scalars fields on the AdS background have been the prototype of
system to study and test the AdS/CFT correspondence. We would like
to show in this section that variational methods provides strong
simplifications in the calculations.

We consider the well known case of a single scalar and reproduce
some well-known results. The action we consider is
\begin{eqnarray}\label{phiaction}
  I &=& \int d^{d+1}x \, \sqrt{\hat g} \left( \frac{1}{2}\hat g^{\mu\nu}\partial_{\mu}\Phi \partial_{\nu}\Phi+
  \frac{1}{2} m^2\, \Phi^2 + \right)
\end{eqnarray}
This example has been extensively analyzed in the literature.
Early calculations \cite{Witten,Freedman-} did not provide the
correct normalization for the vev. This problem was fixed in
\cite{Klebanov-Witten,Muck-Viswanathan,HR}.

The field mass becomes related to the conformal dimension $\Delta$
of the dual operator by \cite{Witten} $m^2 = \Delta(d-\Delta)$. We
follow \cite{HR} and consider in this paragraph operators with
dimensions
\begin{equation}\label{m}
\Delta = {d \over 2} + N, \ \ \ N = 0,1,2,...
\end{equation}
For these masses the asymptotic solution has the Frobenius form with,
\begin{eqnarray}\label{FGs}
\Phi(\rho,x) &=& \rho^{{d-\Delta \over 2}} \varphi(\rho,x)  \label{fro}\\
 \varphi(\rho,x) &=& \varphi_0 + \rho\varphi_1 + \cdots + \rho^{N-1} \varphi_{N-1} + \rho^N (\varphi_N + \psi \log\rho) + \cdots \label{fser}
\end{eqnarray}
The coefficients $\varphi_1,...., \varphi_{N-1}$ and $\psi$ are
determined in terms of $\varphi_0$ by using the asymptotic
equations \cite{HR}. $\varphi_N$ is on the hand undetermined, and
is fixed by imposing boundary conditions at $\rho\rightarrow
\infty$.

Our goal here is to provide a simple proof of the well-known result,
\begin{equation}\label{mas}
\langle {\cal O}  \rangle = (2\Delta -d) \varphi_N + \mbox{local terms},
\end{equation}
using the variational method.

To compute the vev of the dual operator, we vary on-shell the
action and plug the series. As shown in Sec. \ref{renor} we do not
need to worry about divergent terms since they can always be
renormalized away. The variation of the action is
\begin{eqnarray}
  \delta I &=& \int d^dx\,{2 \over \rho^{{d \over 2}-1}} \Phi'\delta \Phi \nonumber\\
   &=&  2\int d^dx\, {1 \over \rho^{N-1}}\, \varphi' \delta  \varphi
\end{eqnarray}
where we have used (\ref{m},\ref{fro}), and discarded a term $\delta (\varphi^2)$. Note that this equation shows clearly why it is necessary to solve the asymptotic equations to order $N$. Now we insert the series (\ref{fser}) and find
\begin{eqnarray}
\delta I &=& 2 \int d^dx {1 \over \rho^{N-1}} \left[\sum_n (n\rho^{n-1} \varphi_n)  + \rho^{k-1} \psi\right]\, \sum_m \rho^m \delta \varphi_m \nonumber\\
&=& 2 \sum_{n,m} \int d^dx \rho^{-N+m+n}\, n \varphi_n \, \delta \varphi_m +   \sum_m\int d^dx\, \rho^m\, \psi \delta \varphi_m \label{sc}
\end{eqnarray}
In the first line we have already discarded $\log$ terms (which remain after computing the derivative in $\varphi'$) since they cannot contribute to the finite vev; they can be either zero, or divergent in which case they are canceled by a counterterm $B(\rho)$.

At this point comes the main simplification of our method. The sum in  (\ref{sc}) contains $\rho-$dependent terms, and a zero mode term. As explained in Sec. \ref{renor} we only need to keep the zero mode. Thus we set $m=k-n$ and find
\begin{eqnarray}
\delta I &=& 2 \sum_{n} \int d^dx \, n \,\varphi_n \, \delta \varphi_{k-n} + \psi \delta \varphi_0\nonumber\\
&=& 2 \int d^dx \left[ \varphi_1 \delta\varphi_{N-1} + 2\varphi_2 \delta\varphi_{N-2} + \cdots + (N-1)\varphi_{N-1} \delta\varphi_{1} +  (N\varphi_N + \psi)\delta\varphi_{0} \right]
\end{eqnarray}
The first $N-1$ terms are purely local  because
$\varphi_1,...,\varphi_{N-1}$ are fixed by the asymptotic
equations as functions of $\varphi_0$.  They can thus be written
in the form $X\delta\varphi_0$ where $X$ is a local function of
$\varphi_0$.  The last term, on the other hand, contains
$\varphi_N$ which is non-trivial. Reinserting $N$ in terms of $d$
and $\Delta$ gives the vev (\ref{mas}), as promised.

\subsection{Coulomb branch flow} \label{CBflow}

In this paragraph we consider another known example \cite{HR},
namely, gravity coupled to a scalar through the action
\begin{equation}\label{I CBf}
 I = \frac{1}{4}\int d^5x \sqrt{\hat{g}}(\hat{R} - 2\Lambda
 ) -\frac{1}{2} \int d^4x \sqrt{h} K + \frac{1}{2} \int d^5x
 \sqrt{\hat{g}}( \hat{g}^{\mu \nu} \partial _{\mu} \Phi \partial _{\nu} \Phi + 2 V(\Phi)
 ) + B + B_0
\end{equation}
\noindent where $V$ is defined by:
\begin{equation}\label{V CBf}
    V(\Phi)  = -2\Phi^2 + \frac{4}{3\sqrt{6}}\Phi^3
\end{equation}
Since $m^2=-4$, the operator dual to $\Phi$ is a scalar operator
$O$ whose conformal dimension is $\Delta=2$.

The boundary terms $B(\rho)$ and $B_0$ must be chosen such that
\begin{equation}\label{def T and O}
    \delta I = \int d^4x \sqrt{g_0} \left[ \frac{1}{2} \langle T^{ij} \rangle \delta g_{0ij}
    + \langle O \rangle \delta \phi_0 \right]
\end{equation}
The bulk metric and the scalar take the form \cite{HR},
\begin{equation}\label{FGE}
    ds^2 = \frac{d\rho^2}{4\rho^2} + \frac{1}{\rho} g_{ij} dx^i
    dx^j ~~~~~~~~~~ g_{ij} = g_{0 ij} + \rho g_{1 ij} + \rho^2 ( g_{2 ij} + \log
    \rho h_{1 ij} + (\log \rho)^2 h_{2 ij} ) + \ldots
\end{equation}
\begin{equation}\label{Phi series CBf}
    \Phi = \rho ( \tilde{\phi}_0 + \log \rho \phi_0 )  + \rho ^2 (
    \tilde{\phi} _1 + \log \rho \phi_1 + (\log \rho)^2 \psi ) +
    \ldots
\end{equation}
The equations of motion yield,
\begin{equation}\label{sol g CBf}
g_{1 ij} = -\frac{1}{2} R_{0 ij} + \frac{1}{12} R_0 g_{0 ij}
~~~~~~~~ h_{1 ij} = h_{ij} + \frac{2}{3}
\phi_0\tilde{\phi_0}g_{0ij}
\end{equation}
\begin{equation}
\nonumber
h_{2 ij} = \frac{1}{3}\phi_0^2 g_{0 ij} ~~~~~~~~ Tr g_2
= \frac{1}{4} Tr(g_1^2) + \frac{2}{3}( \phi_0^2 + 2
\tilde{\phi}_0^2)
\end{equation}
\begin{eqnarray}\label{sol phi CBf}
  \phi_1 &=& -\frac{1}{4}(\Box_0 \phi_0 + \frac{1}{3}R_0\phi_0 ) -
             \frac{4}{\sqrt{6}}(\phi_0^2-\frac{1}{2}\phi_0\tilde{\phi}_0)   \\
\nonumber
  \tilde{\phi}_1 &=& -\frac{1}{4} \left[ \Box_0 \phi_0 +\frac{1}{3}R_0(\phi_0 + \tilde{\phi}_0)
                     + 8 (\phi_1 + \psi) \right] + \frac{1}{\sqrt{6}}\tilde{\phi}_0^2 \\
\nonumber
  \psi &=& \frac{1}{\sqrt{6}}\phi_0^2
\end{eqnarray}
where $h_{ij}$ is the value of the log coefficient for the pure
gravitational case.  Dropping all total variations and
divergencies the on-shell variation of the action reads:
\begin{equation}\label{dI CBf gral}
    \delta I = -\int d^4x \sqrt{g} \frac{1}{\rho} ( k^{ij} -
    kg^{ij}) \delta g_{ij} + 2 \int d^4x \sqrt{g} \frac{1}{\rho}\Phi'\delta \Phi +
    \delta B + \delta B_0
\end{equation}
where $k_{ij}=g'_{ij}$. Plugging the series (\ref{sol g CBf}) and
(\ref{sol phi CBf}) in (\ref{dI CBf gral}) and choosing properly
the boundary terms $B$ and $B_0$, is straightforward to find:
\begin{eqnarray}\label{Tij CBf}
  \langle T^{ij} \rangle  &=& t^{ij} +  \left [ \frac{2}{3}\phi_0^2 + \frac{1}{3}\tilde{\phi}_0^2 -\phi_0\tilde{\phi}_0 \right ] g_0^{ij}, \ \ \ \ \   \langle O \rangle = - 2 \tilde{\phi}_0
\end{eqnarray}
where $t^{ij}$ is the purely gravitational tensor \cite{HR}. These
results, of course, agree with those found in \cite{HR}.

\subsection{Gravity and integrability conditions: Einstein equations}

\label{gravity}

In this section we will consider the case of Gravitation with
negative cosmological constant in $d=4$. The explicit formula for
the energy momentum tensor is known \cite{HR}, and it has also
been considered from the variational point of view in \cite{BST}.
Our goal in this paragraph is to use this system to test the
integrability condition (\ref{dL}). The bulk metric takes the form
\cite{HR},
\begin{equation}\label{FGE}
    ds^2 = \frac{d\rho^2}{4\rho^2} + \frac{1}{\rho} g_{ij} dx^i
    dx^j ~~~~~~~~~~ g_{ij} = g_{(0)ij} + \rho g_{(1)ij} + \rho^2(g_{(2)ij} + \log \rho\,
    h_{ij}) + \ldots
\end{equation}
The values of $g_{(1)ij}$, $h_{ij}$, Tr$g_{(2)ij}$ and Div$\,g_{(2)}$ are local functions of $g_{(0)ij}$. In particular,
\begin{eqnarray}\label{g1}
  g_{(1)ij} &=& \frac{1}{12}R_{(0)} g_{(0)ij} - \frac{1}{2} R_{(0)ij}.
\end{eqnarray}
This value is universal and valid for any gravitational action that accepts an AdS background \cite{Imbimbo}.

In this paragraph we only want to address the divergencies in the variation of the action (for the finite part see \cite{BST}). After subtracting total variations, there remains two divergent terms,
\begin{eqnarray*}
  \delta I_{div} = - \int \sqrt{g{(0)}} \left[ \frac{1}{\rho} \left( g_{(1)}^{ij} - \mbox{Tr}(g_1)g_{(0)}^{ij}
  \right) + \frac{1}{2} \log \rho\, h^{ij}  \right]\delta g_{(0)ij} .
\end{eqnarray*}
According to the general discussion in Sec. \ref{renor}, these terms can be written as total variations. In fact, it is known \cite{HR} that $h^{ij}$  can be written as a total ``divergency",
\begin{equation}
h^{ij} = {\delta \over \delta g_{(0)ij}} \int {\cal A}
\end{equation}
where ${\cal A}$ is the anomaly. This shows that the $\log$ divergency can in fact be written as a total variation.  On the other hand, from (\ref{g1}) it is direct to see that $( g_{(1)}^{ij} - \mbox{Tr}(g_{(1)})g_{(0)}^{ij} )$ is
proportional to the Einstein tensor, and thus, the $1/\rho$ divergency can also be written as a total variation $\delta (\sqrt{g}R)$. These counterterms were first derived in \cite{Balasubramanian-Kraus} by demanding finiteness of the energy-momentum tensor.

\section{Higher order theories: $L= L(q,q',q'',...)$}
\label{HOrder}

We have shown in this paper that for any Lagrangian  $L(\phi,\phi')$ the divergencies appearing in the variation of the action can be canceled by a boundary term.  We would like to show in this paragraph that the same conclusions follow with higher order theories with Lagrangians of the form $L(\phi,\phi'\phi'',...)$. This means that our results are applicable to string theories with higher $\alpha'$ corrections.

Let us work out here the details for a fourth order theory with $L(\phi,\phi',\phi'')$. The variation of the action yields,
\begin{equation}
\delta I = \int \left[\left({\partial L \over\partial \phi''} \right)'' - \left({\partial L \over\partial \phi'} \right)' + {\partial L \over\partial \phi} \right]\delta \phi + \left[\underbrace{ {\partial L \over\partial \phi'}\delta\phi + {\partial L \over\partial \phi''}\delta\phi' - \left( {L \over \phi''} \right)' \delta\phi} \right]_{\rho=\epsilon}
\end{equation}
The first parenthesis gives the Euler-Lagrange equations while the
second is a boundary term.  We shall prove now that the
$\rho-$dependent parts of this boundary term can be written as a
total variation, and hence subtracted from the action. Let us
denote the piece enclosed by the underbrace as $B$. We proceed
exactly as in the second order case. Computing the radial
derivative of $B$ and using the equations of motion we find,
\begin{eqnarray}
 B' &=& \left({\partial L \over\partial \phi'}\right)' \delta \phi+ {\partial L \over\partial  \phi'}\delta\phi' + {\partial L \over\partial \phi''}
\delta \phi'' - \left( {\partial L \over\partial \phi''} \right)'' \delta \phi \nonumber\\
&=& {\partial L \over\partial \phi} \delta \phi + {\partial L \over\partial \phi'}\delta \phi' + {\partial L \over\partial \phi''}\delta \phi'' \nonumber\\
&=& \delta L.
\end{eqnarray}
As in the second order theory, this shows that all divergencies present in $B$ can be renormalized away. The zero mode of $B$ is not restricted by the equation. For higher order theories the boundary data has a bigger structure (more initial conditions) and hence the AdS/CFT dictionary may need extra revision.

\section{Acknowledgments}

We would like to thank S. Theisen, M. Henneaux and C. Nu\~nez for
their useful comments and discussions. We also thank M. Berg, M.
Haack and W. M\"{u}ck for useful comments on an earlier version of
this manuscript. M.B. work is partially supported by FONDECYT \#
1060648.


\begin{thebibliography}{10}

\bibitem{Maldacena} J. Maldacena, Adv. Theor. Math. Phys. {\bf 2},
231 (1998) [arXiv:hep-th/9711200].

\bibitem{GKP} S. Gubser, I. Klebanov, A. Polyakov, Phys. Lett. B
{\bf428}, 105 (1998) [arXiv:hep-th/9802109].

\bibitem{Witten} E. Witten, Adv.Theor.Math.Phys {\bf 2}, 253 (1998) [arXiv:hep-th/9802150].

\bibitem{Klebanov-Witten} I. R. Klebanov, E. Witten,
Nucl.Phys.B {\bf 556}, 89 (1999) [arXiv:hep-th/9905104].

\bibitem{Witten01}
  E.~Witten, [arXiv:hep-th/0112258].

\bibitem{Henneaux} M. Henneaux, [arXiv:hep-th/9902137].

\bibitem{Henningson-Sfetsos} M. Henningson, K. Sfetsos,
Phys.Lett.B {\bf 431}, 63 (1998) [arXiv:hep-th/9803251].

\bibitem{Freedman-}
D.~Z.~Freedman, S.~D.~Mathur, A.~Matusis and L.~Rastelli,
  Nucl.\ Phys.\ B {\bf 546}, 96 (1999)
  [arXiv:hep-th/9804058].

\bibitem{Muck-Viswanathan} W.~M\"{u}ck and K.~S.~Viswanathan,
  Phys.\ Rev.\ D {\bf 58}, 106006 (1998)
  [arXiv:hep-th/9805145].
  W.~M\"{u}ck and K.~S.~Viswanathan, Phys.\ Rev.\ D {\bf 58}, 041901 (1998)
  [arXiv:hep-th/9804035].

\bibitem{HR} S. de Haro, S. Solodukhin, K. Skenderis, Commun.Math.Phys {\bf 217}, 595 (2001) [arXiv:hep-th/0002230]. M.~Bianchi, D.~Z.~Freedman and K.~Skenderis,
  JHEP {\bf 0108}, 041 (2001)   [arXiv:hep-th/0105276].
M. Bianchi, D. Z. Freedman, K. Skenderis,  Nucl.Phys.B {\bf 631},
159 (2002) [arXiv:hep-th/0112119].

\bibitem{Arutyunov-Frolov}
  G.~E.~Arutyunov and S.~A.~Frolov,
  Nucl.\ Phys.\ B {\bf 544}, 576 (1999)
  [arXiv:hep-th/9806216].

\bibitem{Arutyunov-Frolov-Theisen}
  G.~Arutyunov, S.~Frolov and S.~Theisen,
  Phys.\ Lett.\ B {\bf 484}, 295 (2000)
  [arXiv:hep-th/0003116].

\bibitem{Minces-Riveles}
 P.~Minces and V.~O.~Rivelles,
  JHEP {\bf 0112}, 010 (2001)
  [arXiv:hep-th/0110189].
  P.~Minces and V.~O.~Rivelles,
  Nucl.\ Phys.\ B {\bf 572}, 651 (2000)
  [arXiv:hep-th/9907079].

\bibitem{Braga-Boschi}
H.~Boschi-Filho and N.~R.~F.~Braga,
  Phys.\ Lett.\ B {\bf 471}, 162 (1999)
  [arXiv:hep-th/9910233].

\bibitem{Nolland}   D.~Nolland,
  Phys.\ Lett.\ B {\bf 584}, 192 (2004)
  [arXiv:hep-th/0310169].

\bibitem{Muck}W.~M\"{u}ck, Phys.\ Lett.\ B {\bf 531}, 301 (2002)
  [arXiv:hep-th/0201100].

\bibitem{Hatmann-Rastelli}
T.~Hartman and L.~Rastelli, [arXiv:hep-th/0602106].

\bibitem{Berkooz-Sever-}
M.~Berkooz, A.~Sever and A.~Shomer, JHEP {\bf 0205}, 034 (2002)
  [arXiv:hep-th/0112264].

\bibitem{Papadimitriou}
I.~Papadimitriou,  [arXiv:hep-th/0606038].

\bibitem{Marolf-Ross}
D.~Marolf and S.~Ross, JHEP {\bf 0611}, 085 (2006)
[arXiv:hep-th/0606113].

\bibitem{Minces}
P.~Minces, Phys.\ Rev.\ D {\bf 68}, 024027 (2003)
  [arXiv:hep-th/0201172].

\bibitem{HRnew} I. Papadimitriou, K. Skenderis,
[arXiv:hep-th/0404176].
I. Papadimitriou, K. Skenderis, JHEP {\bf 0410}, 075 (2004)
[arXiv:hep-th/0407071]

\bibitem{PS}
  I.~Papadimitriou and K.~Skenderis,
  JHEP {\bf 0508}, 004 (2005)
  [arXiv:hep-th/0505190].

\bibitem{Imbimbo} C. Imbimbo, A. Schwimmer, S. Theisen, S.
Yankielowicz, Class.Quant.Grav. {\bf 17}, 1129, (2000) [arXiv:hep-th/9910267].

\bibitem{ST1} A. Schwimmer, S. Theisen, JHEP {\bf 0008}, 32
(2000) [arXiv:hep-th/0008082].

\bibitem{ST2} A. Schwimmer, S. Theisen, JHEP {\bf 0310}, 001 (2003)
[arXiv:hep-th/0309064].

\bibitem{deBoer-Verlinde2}
  J.~de Boer, E.~P.~Verlinde and H.~L.~Verlinde,
  JHEP {\bf 0008}, 003 (2000)
  [arXiv:hep-th/9912012].

\bibitem{Henningson-Skenderis} M. Henningson, K. Skenderis, JHEP {\bf 9807}, 023 (1998)
[arXiv:hep-th/9806087].

\bibitem{Berg-Haack-Muck}
  M.~Berg, M.~Haack and W.~M\"{u}ck,
  Nucl.\ Phys.\ B {\bf 736}, 82 (2006)
  [arXiv:hep-th/0507285].

\bibitem{BST} M. Ba\~{n}ados, A. Schwimmer, S. Theisen, JHEP {\bf 0405}, 039, (2004)
[arXiv:hep-th/0404245].

\bibitem{BTOM} M. Ba\~{n}ados, R. Olea, S. Theisen,  JHEP {\bf 0510},
067 (2005) [arXiv:hep-th/0509179]. M. Ba\~{n}ados, O. Miskovic, S.
Theisen, JHEP {\bf 0606}, 025, (2006) [arXiv:hep-th/0604148].

\bibitem{HMTZ}
  J.~Gegenberg, C.~Mart\'inez, R.~Troncoso Phys.\ Rev.\ D {\bf 67}, 084007
  (2003) [arXiv:hep-th/0301190]
  M.~Henneaux, C.~Mart\'inez, R.~Troncoso and J.~Zanelli, Phys.\ Rev.\ D {\bf 70}, 044034 (2004)   [arXiv:hep-th/0404236].
  T.~Hertog and K.~Maeda, JHEP {\bf 0407}, 051 (2004)
  [arXiv:hep-th/0404261].
  M.~Henneaux, C.~Mart\'inez, R.~Troncoso and J.~Zanelli, [arXiv:hep-th/0603185].

\bibitem{BST2}
  M.~Banados, A.~Schwimmer and S.~Theisen, JHEP {\bf 0609}, 058 (2006) [arXiv:hep-th/0604165].

\bibitem{Balasubramanian-Kraus}
  V.~Balasubramanian and P.~Kraus,
  Commun.\ Math.\ Phys.\  {\bf 208}, 413 (1999)
  [arXiv:hep-th/9902121].

\end{thebibliography}
 \end{document}